\documentstyle[epsfig]{aipproc}
\newcommand{\be}{\begin{equation}}
\newcommand{\ee}{\end{equation}}
\newcommand{\ba}{\begin{eqnarray}}
\newcommand{\ea}{\end{eqnarray}}
\newcommand{\Mc}{{\cal M}}
\newcommand{\Ms}{M_{\odot}}

\def\ltsima{$\; \buildrel < \over \sim \;$}
\def\simlt{\lower.5ex\hbox{\ltsima}}
\def\gtsima{$\; \buildrel > \over \sim \;$}
\def\simgt{\lower.5ex\hbox{\gtsima}}

\begin{document}

\title{Deep Surveys of Massive Black Holes with LISA}

\author{Alberto Vecchio}
\address{Max Planck Institut f\"{u}r Gravitationsphysik,
Albert-Einstein-Institut\\
Am M\"{u}hlenberg 1, D-14476 Golm, Germany}

\maketitle

\begin{abstract}
Massive black hole binary systems -- with mass in the range $\sim 10^5\,\Ms$ -
$10^8\,\Ms$ -- are among the most interesting sources for the
Laser Interferometer Space Antenna (LISA); gravitational radiation
emitted during the last year of in-spiral could be detectable with a 
very large ($\sim 10^3$) signal-to-noise ratio for sources at 
cosmological distance. Here we discuss the impact of LISA for astronomy
and cosmology; we review our present understanding of the relevant
issues, and highlight open problems that deserve further investigations.

\end{abstract}

\section*{Introduction}

The Laser Interferometer Space Antenna (LISA)~\cite{LISA} is 
a gravitational wave (GW) observatory in the low-frequency band 
which is currently accessible only through non-dedicated (and low
sensitivity) experiments based on the technique 
of Doppler tracking of interplanetary spacecraft~\cite{EW,BVI99}.
As of this writing, LISA is identified as an ESA 
Cornerstone mission in the Horizon 2000-plus program, but is presently
studied by both ESA and NASA with the view of a joint mission with launching
date 2008-2010. The instrument has an optimal sensitivity in the milli-Hz frequency
range, $h_{\rm rms} \approx 3\times 10^{-22}$ for $f \sim 1$ mHz,
covering the band $\sim 10^{-5}\,{\rm Hz} - 30\,{\rm mHz}$.
It consists of a constellation
of tree drag-free spacecraft placed at the vertices of an ideal equilateral triangle 
with sides of $\simeq 5\times 10^6\,{\rm km}$, forming
a three-arms interferometer~\cite{LISA,Danz_am99}.

The low frequency band is populated by a {\it plethora} of GW sources, that are
out of reach for Earth-based detectors, and could be
easily detectable by LISA~\cite{LISA}: they include 
{\it guaranteed} sources, such as {\it known} galactic short-period binary stars;
neutron stars (NS's) and/or low-to-intermediate mass black holes ($\sim 10\,\Ms - 10^3\,\Ms$) 
falling into a massive companion ($\sim 10^5\,\Ms - 10^8\,\Ms$); massive black hole binary 
systems (MBHB's), with mass in the range $\sim 10^5\,\Ms - 10^8\,\Ms$; 
stochastic backgrounds of primordial origin, and generated by the incoherent
superposition of unresolved binary systems in the Universe.

\begin{figure}[b!] % fig 1
\centerline{\epsfig{file=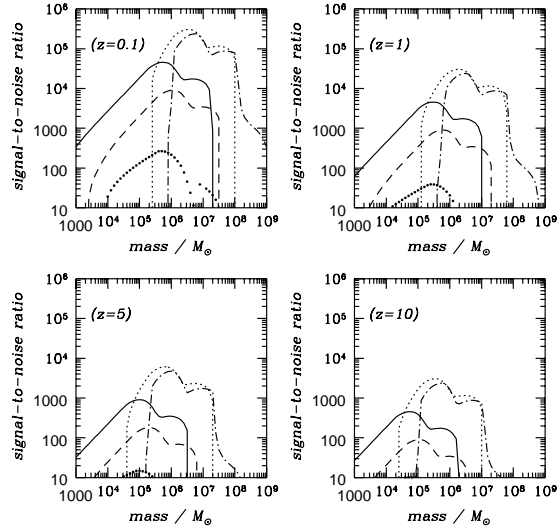,height=3.5in,width=3.5in}}
\vspace{10pt}
\caption{
The sensitivity of LISA to coalescing black hole binary systems. The plots show 
the angle-averaged signal-to-noise ratio which characterizes LISA observations
of the three phases of BH coalescence -- in-spiral,
merger, and ring-down -- as a function of the mass $m_1$ of the primary object.
The solid, dotted and dotted-dashed
lines refer to the in-spiral, merger and ring-down signal, respectively,
of two BH's with $m_1 = m_2$; the dashed line and the bold dots describe the
in-spiral signal from two BH's with $m_1 = 100\,m_2$, and a secondary BH of $10\,\Ms$
orbiting $m_1$, respectively. The SNR of the in-spiral
signal refers to the final year of the source life, with cut-off frequency
$f_{\rm isco} \simeq 4.4\times 10^{-3}\,[m (1 + z)/10^6\,\Ms]^{-1}$; 
the quasi-normal ringing is assumed to occur at $f_{\rm qnr}
\simeq 6.5\times 10^{-2}\,[m (1 + z)/10^6\,\Ms]^{-1}$, and the energy radiated
during the merger and the ring-down phases are computed according to~{\protect\cite{FH98}}.
The four panels refer to different
distances of the fiducial source, where we adopt, for simplicity, a
luminosity distance given by $D = z/75\,{\rm km}\,{\rm sec}^{-1}\,{\rm Mpc}^{-1}$.
The instrument low and high frequency cut-offs are (conservatively)
$10^{-4}$ Hz and $3\times 10^{-2}$ Hz, respectively. The 
noise spectral density takes into account both the instrumental
noise and the so-called confusion noise~{\protect\cite{HBW90,BH97}} .
}
\label{fig:1}
\end{figure}%

The purpose of this contribution is to discuss the impact
of LISA for astronomy. Being impossible to cover all aspects, we 
will concentrate on one specific class of sources:
massive black hole binary systems. We will describe
how LISA works as GW telescope -- we are ultimately dealing with a new
branch of observational astronomy --  summarize 
our present understanding of the main issues,
and highlight open questions that deserve further investigations.

MBHB's are possibly the strongest sources of GW's that LISA
will be able to detect; for typical objects of mass $\sim 10^6\,\Ms$
at redshift $z \sim 1$, the signal-to-noise ratio (SNR) is $\sim 10^3$,
as show in Fig.\ref{fig:1}. The instrument is able to detect
the radiation emitted during one (or more) of the three phases of black hole
coalescence (in the GW jargon: in-spiral, merger, and ring-down)
for a very wide range of masses -- in principle from $\sim 1\,\Ms$ 
to $\sim 10^9\,\Ms$, depending on the mass $m_1$ and $m_2$, 
and the source distance -- see Fig.~\ref{fig:1}, possibly beyond 
redshift $z\sim 5$, if BH's do already exist, and are involved in catastrophic
events with copious release of energy through GW's. 

LISA will be able to carry out a deep and extensive census of black hole 
populations in the Universe, providing an accurate demography of these 
objects and their environment. Compelling arguments suggest
the presence of MBH's in the nuclei of most galaxies, and they are
invoked to explain a number of phenomena, in particular the activity
of quasars and active galactic nuclei~\cite{ZN64,Salpeter64}. However, the observational 
evidences of MBH existence come mainly from observations of relatively
nearby galaxies, whose nuclei do not show significant activity~\cite{Miyoshi95,EG96,maoz}.
Massive black holes seem to be clustered in the mass-range 
$10^6\,\Ms - 10^9\,\Ms$~\cite{Richstoneetal98}; 
at the lower edge of the BH mass-spectrum, we find evidences for solar-mass
BH candidates~\cite{Rees98}. No information is presently available regarding 
BH's with mass between $\sim 10\,\Ms$ and $\sim 10^6\,\Ms$, although
some recent X-ray observations are interpreted as possible (but not compelling) 
indications of "middleweight" BH's~\cite{CM99,PG99}. LISA -- and Earth-based laser 
interferometers -- 
will definitely show whether this gap is simply due to a "selection" effect 
of present electro-magnetic observations, or indeed Nature does not provide
intermediate mass black holes: an important feature of LISA
is its capability of detecting BH's with mass $\sim 10^3\,\Ms - 10^4\,\Ms$,
still far from coalescence at high redshift, see Fig.~\ref{fig:1}. 
LISA is also likely to monitor binary systems 
with a wide spectrum of BH spins and orbital eccentricities, which
will enable us to carry out high precision tests of general 
relativity~\cite{Hughes_am99,Poisson96,Ryan97}, and to derive
a map of the distribution of these physical parameters in astrophysical
objects. 

One of the most interesting observations would be the detection
of GW's {\it and} electro-magnetic radiation from the merger
of two BH's. We do not know as yet, whether a burst of electro-magnetic
radiation is emitted during MBH collisions~\cite{BBR}. Determining 
where and when a MBH merger takes place, and possibly alerting 
in advance the astronomers is of paramount importance; this issue is directly 
linked to the identification of the source host galaxy: it would allow us
to establish correlations between MBH's and their environment, and
use LISA observations to estimate the fundamental cosmological 
parameters~\cite{LISA,Schutz86}.

We have not discussed so far the rate at which we expect to detect 
such signals. A fair
statement would probably be that, essentially, we do not  know it.
However, we can
summarize our present knowledge as follows. For MBHB systems, the event
rate depends strongly on theoretical prejudices and model assumptions;
the "canonical" value is $\sim 1\,{\rm yr}^{-1}$, but rates as high
as $\sim 10^3\,{\rm yr}^{-1}$ or as low as $\sim 10^{-2}\,{\rm yr}^{-1}$ are
consistent with theoretical models~\cite{BBR,Bleas_am99,Hae,Vecchio97}. For low-mass
black holes captured by a massive one in galactic cores, we believe
to have a better understanding, and current astrophysical estimates
yield a rate of a few events per year up to $z\simeq 1$~\cite{SR,Sigurdsson97}.

\section*{The LISA telescope}

We are dealing with a new generation of telescopes, both regarding
the kind of radiation they observe (gravitational waves) and the 
frequency window in which they operate ($\sim$ mHz).
It is therefore instructive to analyze the features that enable LISA
to extract accurate information about GW sources. 

We consider here only the in-spiral portion of the whole coalescence waveform,
neglecting the merger and ring-down, 
both easily detectable, cfr. Fig.~\ref{fig:1}. The merger waveform is still poorly 
understood from the theoretical point of view; significant progresses 
have been made using either full numerical schemes or
semi-analytical approximations, but both approaches are still far from returning a satisfactory
answer for GW observations (see~\cite{GC_web,Pullin} and references therein). 
We do however expect to 
gain key information by detecting GW's emitted during the final plunge, 
for instance how energy and angular momentum are radiated during this extreme 
strong-gravity phase. The ring-down signal, 
on the contrary, is theoretically well know; in order to limit the level of complexity
of our analysis, we do not include it into the signal that we consider here; however,
future investigations should keep it (as well as the final plunge, if/when available)
into account, as it might change (conceivably improve)
LISA performances in a number of astrophysical situations.

There are two main features that distinguish the in-spiral signals
recorded by LISA from the ones that we expect to detect with Earth-based
interferometers: (i) they last for months-to-centuries 
(depending on the masses) in the instrument observational band, and
therefore are not burst-signals; in fact, the (Newtonian) time to coalescence 
is $\tau \simeq 1.2\times 10^7\,\left(f_0/10^{-4}\,{\rm Hz}\right)^{-8/3}\,
\left[m\,(1 + z)/10^6\,\Ms\right]^{-5/3}\,
\left(\eta/0.25\right)^{-1}\,{\rm sec}$;
here $m = m_1 + m_2$ is the total mass, and $\eta = \mu/m$
is the symmetric mass ratio, where $\mu = m_1 m_2/m$ is the reduced mass;
(ii) the structure of the waveform is in general much
more complex; in fact, we can expect to detect
black holes that are fast spinning {\it and} live on highly elliptical orbits, in particular
for the extreme mass ratio case, $\eta \ll 1$~\cite{HB95}. As an example, in LIGO 
observations one will likely monitor no more than 
10 cycles of precession of the orbital plane and the spins, whereas in the LISA band,
for a typical observation time of 1 year, they could be as many as $\sim 1000$, see Table~\ref{tab:1}. %
\begin{table}[b!]
\caption{The number of precession cycles observed by LISA. The table shows the number
of cycles (${\cal N}_{\rm prec}$) of ${\bf \hat L}$ and ${\bf \hat S}$ around the constant direction
of the total angular momentum ${\bf J} = {\bf L} + {\bf S}$
during the final year of in-spiral for BH binary systems with selected masses 
(in units of $\Ms$) and spins.
}
\label{tab:1}
\begin{tabular}{c|cc|cc|cc|cc}
$S/m^2$ & $m_1$ $m_2$ & ${\cal N}_{\rm prec}$  
& $m_1$ $m_2$ & ${\cal N}_{\rm prec}$ 
& $m_1$ $m_2$ & ${\cal N}_{\rm prec}$  & $m_1$ $m_2$ & ${\cal N}_{\rm prec}$\cr
\hline
0.95 & $10^7$   $10^6$  &              11 & $10^6$  $10^6$  &   25&  $10^6$  $10^5$  &   23
& $10^6$  $10^2$  &  1262\cr
0.50 & $10^7$   $10^6$  &               7 & $10^6$  $10^6$  &   20&  $10^6$  $10^5$  &   16
& $10^6$  $10^2$  &   708\cr
0.10 & $10^7$   $10^6$  &               4 & $10^6$  $10^6$  &   16&  $10^6$  $10^5$  &    9
& $10^6$  $10^2$  &   150\cr
0.01 & $10^7$   $10^6$  &               3 & $10^6$  $10^6$  &   16&  $10^6$  $10^5$  &    8
& $10^6$  $10^2$  &    16\cr
\end{tabular}
\end{table}%

An useful figure, for both detection and parameter estimation, is also the 
number of wave cycles recorded by LISA:
during the final year of in-spiral, they range from $\sim 10^3$ 
(for $m_1 \sim m_2$) to $\sim 10^5$ (for $\eta \ll 1$). 

In general, 17 parameters describe the waveform. No analysis has been
carried out so far dealing with such general situation. Here, we will introduce
some simplifying assumption, while retaining most of the key physical 
ingredients. The main limitation of our approach derives from 
considering circular orbits; this is probably quite realistic for binary systems of two
MBH's which have undergone a common evolution inside a galactic core, but is
almost for sure violated for solar mass compact objects and/or low mass 
BH's orbiting a massive one~\cite{HB95}. We do, however, take into account spins; 
in this case we assume that either the masses of the BH's are 
roughly equal, or one of the BH's has a negligible spin (which still describe 
a wide range of astrophysical situations): the binary system undergoes the so-called 
{\it simple precession}~\cite{ACST94}, where the orbital angular momentum
${\bf L}$ and the total spin ${\bf S} = {\bf S_1} + {\bf S_2}$ are locked
together, and precess around the (almost) constant direction of the
total angular momentum ${\bf J}  = {\bf S} + {\bf L}$. We also use
the post$^{1.5}$-Newtonian approximation of the GW phase~\cite{BDIWW}. 
As a consequence of this chain of approximations, the number of parameters 
describing the signal drastically reduces, from 17 to 11. %
\begin{figure}[b!] % fig 1
\centerline{\epsfig{file=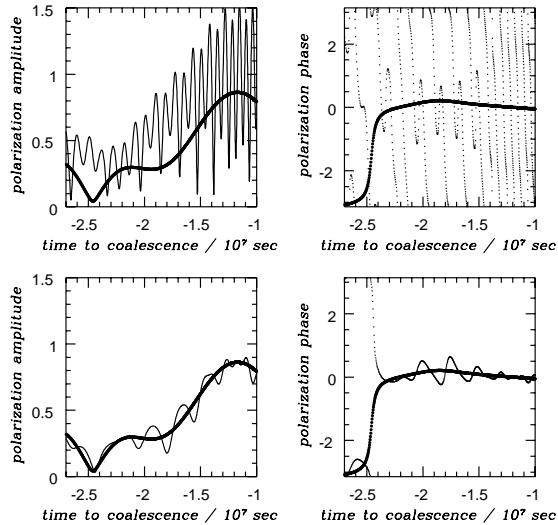,height=3.5in,width=3.5in}}
\vspace{10pt}
\caption{In-spiral signals at the LISA output. The plots show the 
evolution of the polarization amplitude $A_{\rm p}(t)$ (on the left) 
and phase $\varphi_{\rm p}(t)$ (on the right)  as a function of time,
cfr. Eq.~{\protect\ref{h}}, for black holes with $S = 0$
(bold solid line) and $S\ne 0$ (thin solid line, for $A_{\rm p}(t)$,
and dotted line, for $\varphi_{\rm p}(t)$).
The two plots at the top refer to  
a source with masses $m_1 = 10^7\,\Ms$ and $m_2 = 10^5\,\Ms$; in the case
of spinning black holes the parameters are:
$S/m^2 = 0.95$, ${\bf \hat S} \cdot {\bf \hat L} =0.5$. The plots at the
bottom refer to a MBHB with $m_1 = m_2 = 10^6\,\Ms$: when spins are
present the choice of parameters is according to: $S/m^2 = 0.3$, and
${\bf \hat S} \cdot {\bf \hat L} =0.9$. 
}
\label{fig:2}
\end{figure}%

It is useful now to review some of the instrumental features, in order to understand
how LISA works as GW observatory: 

(i) LISA is an {\it all-sky monitor}, and one gets for free
all-sky surveys. During the observation time, however, LISA changes
location and orientation. The LISA orbital motion is rather peculiar
-- the baricenter of the instrument is inserted in a heliocentric orbit, following
by $20^o$ the Earth; the detector plane is tilted by $60^o$ with respect to
the Ecliptic and the instrument counter-rotates around the normal to the 
detector plane with the same 1-yr period -- and
is conceived in order to keep the configuration as stable as possible during the mission, 
as well as to give optimal coverage of the sky. It also turns out to be a key
factor in reconstructing the source location
in the sky. 

(ii) The sources are distinguished in the data stream by the different 
structure and time evolution of the signals at the detector output; the recorded
in-spiral signal reads:
\be
h_{\alpha}(t) = A_{\rm gw}(t)\,A_{\rm p}^{\alpha}(t)\,\cos[\phi_{\rm gw}(t) +
\varphi_{\rm p}^{\alpha}(t) + \phi_{\rm D}(t)]
\label{h}
\ee
where $A_{\rm p}(t)$ and $\varphi_{\rm p}(t)$ are the time-varying polarization amplitude
and phase, respectively, and $\phi_{\rm D}(t)$ is the Doppler phase shift
induced by the motion of the detector around the Sun; 
an example of in-spiral signal at the output of LISA is given in Fig.~\ref{fig:2}.
The signal is therefore
amplitude and phase modulated by the motion of the LISA centre-of-mass around the Sun,
the change of orientation of the detector arms, and of the
binary orbital plane. All these effects encode
information about some of the source parameters. 

(iii) There is only one LISA detector 
currently planed; correlations and/or time-of-flight measurements are not possible;
they would be highly desirable 
in order to improve the estimation of the source parameters, in particular
the source location and distance; however, as
the gravitational wavelength is $\lambda_{\rm gw} \simeq 2\,(f/1\,{\rm mHz})^{-1}$ AU,
a second detector would have to be placed at several AU from the first one
in order to provide useful information on the position of a source in the sky; however
LISA is a three-arms instrument; Cutler~\cite{Cutler98} has shown that the outputs 
from each arm can be combined
in such a way to form a pair of data sets, $\alpha = 1,2$ in Eq.~(\ref{h}),
whose noise is uncorrelated at all frequencies, that are equivalent
to the data streams recorded by two co-located interferometers,
rotated by $\pi/4$ one with respect to the other. 

Indeed, there will be two data streams available
to extract all source parameters. Correlations between the parameters are 
inevitable, and conspire to degrade the accuracy of the parameter measurements. 
It should also be clear that for LISA
the measurement errors depend crucially on the actual value of the source parameters,
and one therefore needs to explore a very
large parameter space to give a fair description of the instrument performances. %
\begin{figure}[b!] % 
\centerline{\epsfig{file=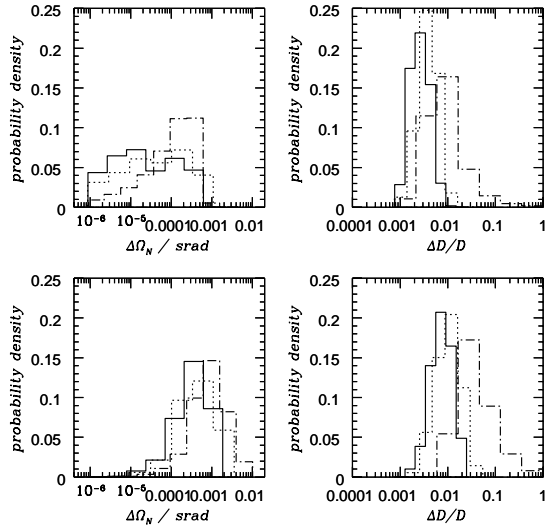,height=3.5in,width=3.5in}}
\vspace{10pt}
\caption{The probability distribution of the angular resolution $\Delta\Omega_N$
and the relative error of the distance determination $\Delta D/D$, 
with which LISA can identify a MBHB by observing the final year of in-spiral. 
The histograms show the result of a Monte-Carlo simulation, where 1000
sources, with masses $m_1 = m_2 = 10^6\,\Ms$ at redshift $z=1$, have  
been randomly located and oriented in the sky.
The top panels refer
to measurements carried out by both LISA detectors, whereas the
bottom panels report the results obtained by using only a single
interferometer. The plots compare the estimated errors in the measurement
of the parameters assuming three different values of the spin:
$S/m^2 = 0.9$ (solid line), 0.3 (dotted line), and 0
(dotted-dashed line). The total noise is given by the sum of the instrumental
noise and the confusion noise.
}
\label{fig:3}
\end{figure}%

\section*{Surveys of massive back holes}

We have discussed in the Introduction the sensitivity of LISA: 
there is little doubt that such interferometer 
will be able to survey a fairly large fraction BH populations in the Universe. 
We would like to stress that in the present discussion, we assume to be able to 
monitor the whole final year of in-spiral. This is a key and delicate point 
which affects the capability of surveying sources at increasingly higher $z$ 
and/or with larger $m$, and measuring precisely the parameters: in fact, 
at some frequency (between $10^{-4}$ Hz and $10^{-5}$ Hz) the instrumental noise 
will completely dominate the signal, allowing to pick up only the very final
portion of the in-spiral (say a few days), or even preventing the
detection; the redshifted radiation simply falls outside the observational band,
cfr. Fig.~\ref{fig:1}. 
It is clear that the higher 
the redshift, the lower the typical mass for which LISA reaches the optimal sensitivity.
Super-massive black holes of mass $\sim 10^9\,\Ms$ might
be observable, by detecting ring-down signals at low redshifts ($z\simlt 0.1$),
if the sensitivity window extends to $\sim 10^{-5}$ Hz. 

Several analysis have been carried out so far dealing with the accuracy of
the parameter measurements with LISA~\cite{Cutler98,VC98,Vecchio99,Sintes_am99,HM99}; 
however, they have been mainly focussed on investigations of the instrument 
angular resolution; moreover, spin effects have been either ignored or 
explored for a very limited portion of the total parameter range. Here we will
try to give a more comprehensive description of the performances of LISA as 
GW observatory.
The accuracy of the parameter measurements is very sensitive to 
the actual source parameter values; it is therefore almost impossible to give
{\it typical figures} for LISA as GW telescope, that can be applied to
a wide range of binary systems. We discuss
in some detail the case of an equal-mass MBHB, with $m_1 = m_2 = 10^6\,\Ms$,
and give some general criteria to extend these results to other
parameter values. It turns out that the source location and orientation with
respect to the detector play a key role. We have therefore 
performed Monte-Carlo simulations, where we fix the source distance
and the physical parameters, and vary randomly the "geometrical" parameters, 
${\bf \hat N}$, ${\bf \hat J}$ and ${\bf \hat S}$. %
\begin{figure}[b!] % 
\centerline{\epsfig{file=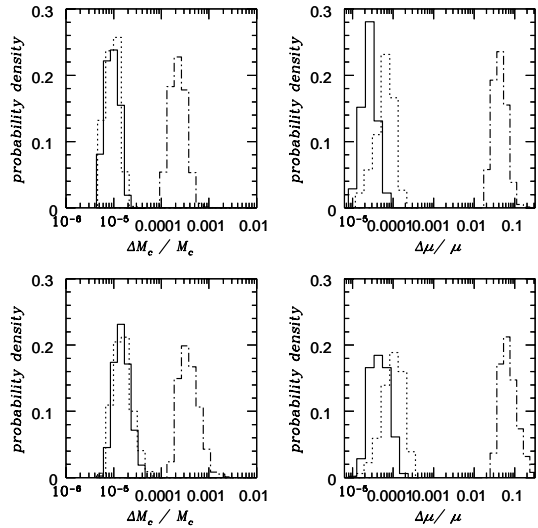,height=3.5in,width=3.5in}}
\vspace{10pt}
\caption{The probability distribution of the errors with which the source masses
can be measured by LISA in one year of observation.
Same as Fig. 3, but now the histograms show the distribution of the 
errors regarding the mass parameters, where our choice corresponds to
the chirp mass, $\Delta M_c/M_c$ (panels on the left), and the reduced 
mass, $\Delta \mu/\mu$ (panels on the right).
}
\label{fig:4}
\end{figure}%
We compute the estimated mean squared errors
associated to the parameter measurements by means of the so-called 
{\it variance-covariance} matrix~\cite{CF94,NV98}.

The main results are presented in Figs.~\ref{fig:3} and~\ref{fig:4}, 
and can be summarized as follows. The angular resolution is $\Delta\Omega_N\sim 10^{-5}$ 
srad; however, depending on the location and orientation of the source 
it varies over a wide range of values, $10 \,{\rm arcmin}^2 \simlt \Delta\Omega_N
\simlt 3\,{\rm deg}^2$. Typically, large spins and misalignment angles -- the angle
between ${\bf \hat L}$ and ${\bf \hat S}$ --
allow us to measure more precisely the source location; for a small region of these
parameters, the "error-box" in the sky could possibly be only a fraction of 
${\rm arcmin}^2$. The distance is usually measured with an error 
$0.1\% \simlt \Delta D/D\simlt 1\%$. The timing accuracy is very high,
and the instance of coalescence can be identified within $\sim 10$ sec. Masses
and spins can be measured very precisely; typically, the errors
affecting the determination of the chirp and reduced mass are
$\Delta \Mc/\Mc \sim 10^{-5}$ and $\Delta \mu/\mu \sim 10^{-4}$,
respectively; the so-called spin-orbit parameter $\beta$ can be determined
with an error $\Delta \beta \sim 10^{-3}$. There is one general rule that can be derived
from this analysis: if BH's are highly spinning and the misalignment angle
is large, the parameter determination improves.
This is due to the fact that the parameters leave peculiar finger
prints on the recorded signal, cfr. Fig.~\ref{fig:2}: in particular, $A_{\rm p}$ and
$\varphi_{\rm p}$ undergo strong modulations, which carry information
not only on the position of the source and the orientation of the angular momenta,
but also on the physical parameters, such as the masses. This is an effect which
is similar -- although the physics behind it is different -- to the one that takes 
place when spins are not present, but one considers
not only radiation emitted at twice the orbital frequency, but also at
other harmonics~\cite{Sintes_am99} (notice that in Fig.~\ref{fig:3} and~\ref{fig:4},
for the case $S = 0$, we report results obtained considering only the dominant
harmonic; we refer the reader to~\cite{Sintes_am99} for more details). 

We can now ask how these results change by selecting different 
source parameters. MBHB's with $m_1\sim m_2\sim 10^7\,\Ms$ would be typically 
observed with larger errors, by a factor $\approx 10$, than the ones reported
here. If we fix $m_1$ and vary $m_2$, the measurement accuracy is fairly constant
-- within, say, a factor $\approx 2$ --
as long as $m_2/m_1 \simgt 0.1$, then is starts degrading: this is due to a rather complex 
competition between several effects, in particular the SNR and the number
of wave/precession cycles~\cite{VCS,AV}. 

MBHB's will be visible several months before the
final coalescence. This will allow us to pick up the signal when the binary system
is still far from merging,
and refine the source parameter measurements as the source proceeds toward the
deadly plunge~\cite{VCS}: for a limited region of the parameter space,
it could be possible to determine the source location in the sky with enough
precision to have a realistic chance of observing the same field with 
other telescopes.

\end{document}